\documentclass[a4paper]{article}

\usepackage[english]{babel}
\usepackage[utf8x]{inputenc}
\usepackage[T1]{fontenc}
\usepackage[parfill]{parskip}
\usepackage{verbatim}
\usepackage{amsfonts}

\usepackage[a4paper,top=3cm,bottom=2cm,left=3cm,right=3cm,marginparwidth=1.75cm]{geometry}

\usepackage{amsmath}
\usepackage{graphicx}
\usepackage[colorinlistoftodos]{todonotes}
\usepackage[colorlinks=true, allcolors=blue]{hyperref}

\title{Categorical Data Integration for Computational Science}
\author{Kristopher Brown, David I. Spivak, Ryan Wisnesky}

\begin{document}
\maketitle

\begin{abstract}
Categorical Query Language is an open-source query and data integration scripting language that can be applied to common challenges in the field of computational science. We discuss how the structure-preserving nature of CQL data migrations protect those who publicly share data from the misinterpretation of their data. Likewise, this feature of CQL migrations allows those who draw from public data sources to be sure only data which meets their specification will actually be transferred. We argue some open problems in the field of data sharing in computational science are addressable by working within this paradigm of functorial data migration. We demonstrate these tools by integrating data from the Open Quantum Materials Database with some alternative materials databases. 
\end{abstract}


\section{Background}

\subsection{Categorical Query Language}
Categorical Query Language (CQL) is based on an algebraic database formalism motivated by category theory \cite{algebraic_databases}, which describes databases as sets of equations, distinct from the other data models (relational, RDF, graph, LINQ).  This formalism is based on a principled theory for data integration and, in particular, facilitates static guarantees of data integrity enforcement using a rich language for constraints\footnote{This property is important for practical applications because the general problem of checking if an arbitrary schema mapping preserves constraints is undecidable and requires an automated theorem prover.}, provides complete provenance for all transformations, and seamlessly integrates programming languages into the schema. \textit{Data integration} will be taken to mean the combination of two database instances with different schemas into a coherent unified schema and instance. The migration of data from one schema into another and even the simple and common act of querying a given database are specific cases of this \cite{algebraic_data_integration}.

Algebraic databases have been used to integrate financial data, health records, and manufacturing service databases\cite{Manufacturing}. Furthermore, the representation of materials science knowledge within category-theoretic data structures has enabled novel analyses of materials \cite{protein,prototype}. These are particularly amenable to describing connections between first principles and emergent properties in systems with many hierarchical scales (other disciplines that have benefited from these tools include physics, linguistics, and computer science \cite{category}).

In this paper, we consider the application of algebraic data integration to computational science more broadly.

\subsection{Data Sharing}

In a 2015 Viewpoint on data sharing\cite{Kitchin}, Kitchin identifies tables and figures in manuscripts to be the current standard for communicating data in scientific research. The flexibility of this medium is achieved at the cost of machine readability and re-usability. \textit{Data} is too broad of a concept to allow for a satisfactory centralized database, thus in most circumstances researchers who share \textit{structured} data must independently release it in some personalized format.

Heterogeneity in data formats leads to challenges, however. A review on academic data sharing practices revealed that a large concern that prevents sharing is the fear of data being misinterpreted and the work needed to address questions and concerns of the data receivers. Furthermore the reliance of unstructured metadata was also cited as a problem that hindered the usability of shared data \cite{drives}. Guidelines have been called for to ease this process: a 2016 review of data sharing in computational science \cite{impact} outlined the field's future challenges: addressing the lack of structured data, tools to combine information from disparate datasets, and  communication of the nuances of data sets.

\begin{figure}[htp]
\centering
\includegraphics[width=13cm]{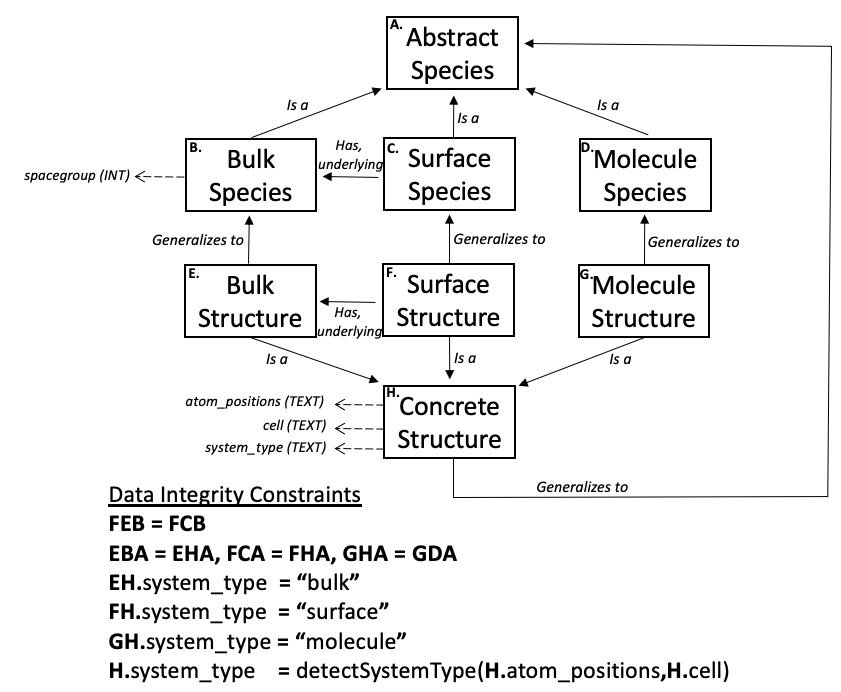}
\caption[LoF Entry]{An CQL schema relating concrete chemical structures and general chemical species. Given a concrete structure, it is possible to classify it as a particular chemical species. We could model this as two entities with one \textit{generalizes to} relation, but we achieve greater granularity by breaking each into subsets which have distinct attributes and relations. The additional \textit{is a} relations are simply the natural injection functions of a subset into some parent set. \\

We can interpret the first integrity constraint as saying "Every surface structure $S$ has the property that: if one looks at the surface species $S'$ that $S$ generalizes to and obtain the underlying bulk species $B'$ of $S'$, and if another finds the bulk structure $B$ that underlies $S$ and obtain the bulk species $B''$ that $B$ generalizes too, then we guarantee that $B' = B''$." Each equation can be written out in this style.\\

The second, third, and fourth constraints enforce our intended meaning of \textit{generalizes to} in spite of the complications posed by explicitly representing bulk, surface, and molecular subclasses. The last four constraints ensure structures are partitioned into bulks, surfaces, or molecules, in accordance with some program \textbf{detectSystemType} that analyzes key structural information.}
\label{olog}
\end{figure}

Although no system safeguards against fundamentally misunderstanding the domain being modeled by some database, data misuse can be mitigated by leveraging the additional features of an CQL schema not present in SQL schemas. CQL schemas are structurally equivalent to ontology logs (ologs) \cite{olog}. Ologs are like "concept maps" that have a formal interpretation. Boxes can be interpreted as database tables with arrows as foreign keys, or alternatively as sets with arrows as functions. They can express nontrivial facts because they are equipped with equations stating when two paths must yield the same result. Thus, an CQL schema is not merely a container for information but itself conveys important conceptual knowledge and can distinguish sensible from nonsensical instance data. As an example, Figure \ref{olog} considers an example from computational materials science, describing the relation between concrete chemical structures (the types of things that are manipulated in computational chemistry, with precise coordinates for atoms) and general chemical species (the kinds of things that appear in chemical reactions, e.g.\ the concept of "a water molecule" or "an FCC-Copper-111 surface").

\begin{figure}[htp]
\centering
\includegraphics[width=12cm]{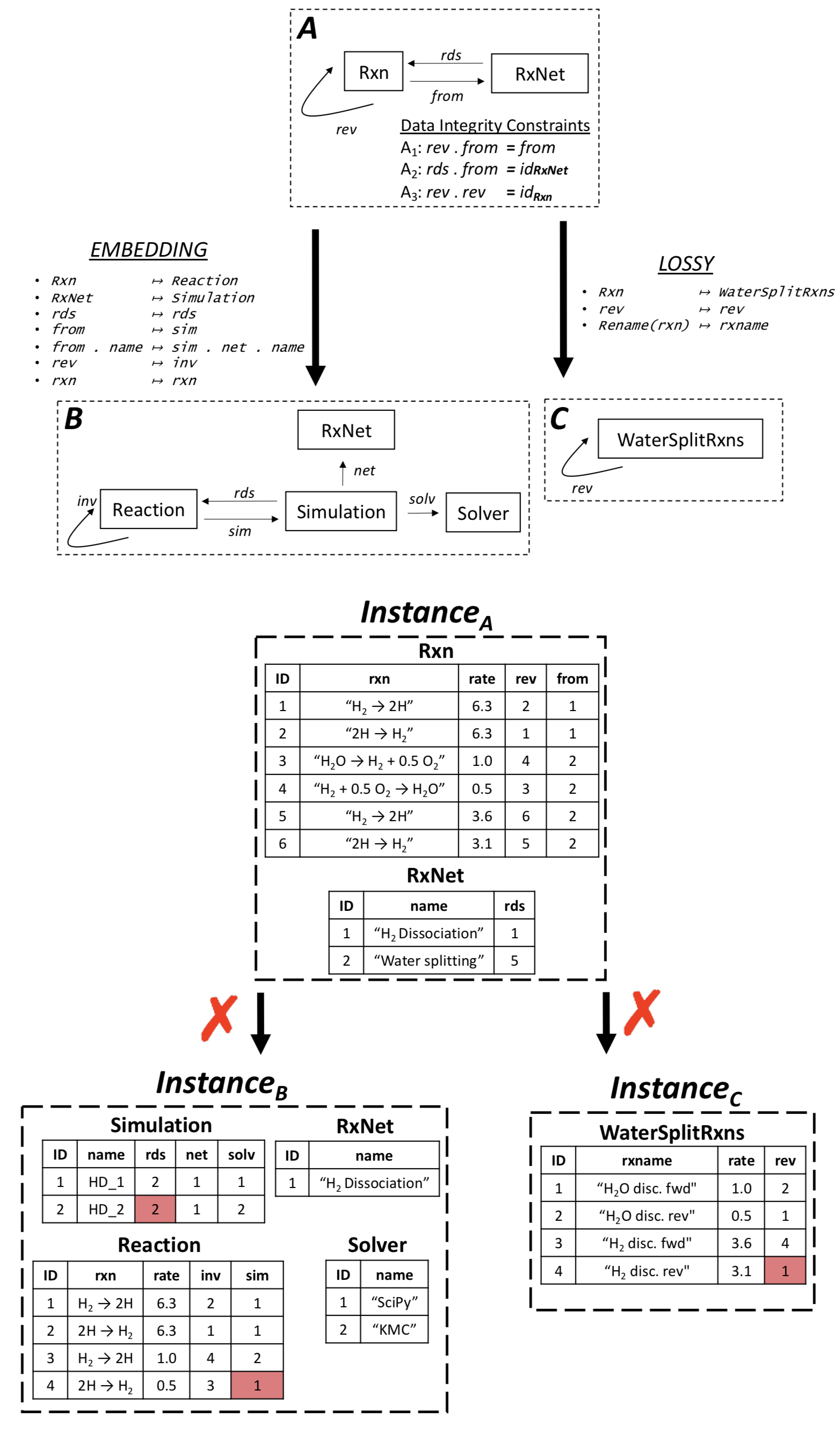}
\caption{\textit{Embedding} and \textit{Lossy} are plausible strategies for migrating data from \textbf{A} into schemas \textbf{B} and \textbf{C}, respectively. Even with these migration strategies in mind, with \textit{ad hoc} data migration scripting it is possible to move the data in a way that, in some sense, does not respect the intent of \textbf{A}. CQL allows us to make this vague notion concrete. We show how the declaration of data integrity constraints ($A_1$, $A_2$, $A_3$) \textit{in the language of \textbf{A}} prevents other researchers from invalidly moving data into their databases through two examples of possible data migrations which are forbidden in the paradigm of functorial data migration. The migration to \textbf{C} is invalid because there exists a reaction (ID = 3) whose reverse's reverse is not the same reaction – we know how to interpret $A_3$ in \textbf{C} because we mapped the \textit{rev} relation in $A$ onto the \textit{rev} relation in $C$.\textsuperscript{*} The migration to \textbf{B} is invalid because there is a simulation (ID = 2) whose \textit{rds} points to a reaction which belongs to a different simulation (violating $A_2$). Furthermore, there exists a reaction (ID = 3) whose inverse does not belong to the same simulation (violating $A_1$).}
\small\textsuperscript{*}CQL prevents these migrations by analyzing the schema mappings, not by checking data after the fact.
\label{rxn}
\end{figure}

These path equations communicate higher-level meaning to those who view the database and also serve as data-integrity constraints for any instance data that is stored in the schema, which helps prevent a large class of data misuse errors in two ways. Firstly, there is a static guarantee that data represented in an algebraic database cannot violate these constraints. Thus, a researcher with a particular frame of reference can safely draw data from external sources without worrying that the new data violates their data integrity constraints. Secondly, CQL transformations come from the paradigm of \textit{functorial data migration} \cite{fdm}, meaning they are structure-preserving in both the source and target schemas. This protects those who \textit{share} data, which can only be migrated in a way that respects the rules that accompany the source schema. Without these abstractions, users must write scripts to perform data migrations by directly manipulating databases with raw SQL, which is error-prone as well as fragile to changes in source or target schema specifications.  Figure \ref{rxn} considers moving data of chemical reaction network simulations between different representations and provides concrete examples of erroneous data migrations which are \textit{not} possible to perform with CQL.

In summary, there exist systemic issues in academic data sharing given the lack of a specific and universal standard for data representation. The formalism of category theory offers a flexible, machine-interpretable language for declaring customized data structures and translating between them such that the intended meanings of the data provider and the data receivers are respected; mediating these transformations with CQL provides assurances of data quality by construction. 

\pagebreak

\subsection{Computational Science}

Simulation now serves as a complementary tool of scientific inquiry, alongside theory and experiment. As a case study, we consider Density Functional Theory (DFT), a popular tool for materials scientists which enables the efficient prediction of electronic properties (such as stability) from a chemical structure input \cite{burke}. Databases of computational data benefit the public in many ways\cite{impact}: 
\begin{enumerate}
    \item to aid and inspire further computational studies
    \item to establish reference against which to compare results
    \item to facilitate property prediction and materials screening
\end{enumerate}

Data sharing can greatly accelerate the rate of research, as high quality data is expensive to generate. However, there are few domains in which widely-accepted standards for data sharing exist. For example, although there are many standardized formats for representing chemical structures \cite{babel}, there are no such standards for more complicated entities such as the symmetry analysis of a structure, the pseudopotentials used in DFT calculation, density of states data resulting from a DFT calculation, and the DFT calculation itself. 

A quantum chemistry dataset can be considered as a container for many DFT calculations. Each calculation can involve multiple structures in addition to having a large number of parameters and metadata; there is no standard format for storing data at this higher level of abstraction. This is compounded by the fact that the meaning of a "calculation" is not something well-defined: even if one fixes the software package which implements DFT and its required input parameters, in practice DFT gets called by arbitrary scripts which do nontrivial sequences of commands as well as on-the-fly data preprocessing and postprocessing. Thus a DFT calculation has no \textit{a priori} restrictions on what its inputs or outputs might be.

Furthermore, many questions of interest depend not on a single calculation, but rather on ensembles of calculations, grouped in particular ways; for example, one is often interested in  formation energies, i.e.\ a structure's energy relative to some reference energy, which depends on some arbitrarily-chosen mapping of its constituent elements to reference species, as well as the calculations for those reference species. For this to be valid, those calculations ought be done at a similar (though usually not exact) DFT settings.

The above descriptions represent a tiny fraction of the complexity of the systems computational scientists grapple with. In practice, scientists can only communicate \textit{structured} raw data in tiny fragments (e.g.\ specific chemical structures) of the systems they try to model, which contain concepts at higher levels of abstraction such as chemical species, reaction mechanisms, and reaction networks. Although not currently feasible, the ability to freely exchange structured data at the level of abstraction which scientists actually work at would support many scientific activities such as machine learning applications which thrive on large datasets, straightforward testing of results for reproducibility, and avoiding duplication of computational effort.

\begin{figure}[htp]
\centering
\includegraphics[width=10cm]{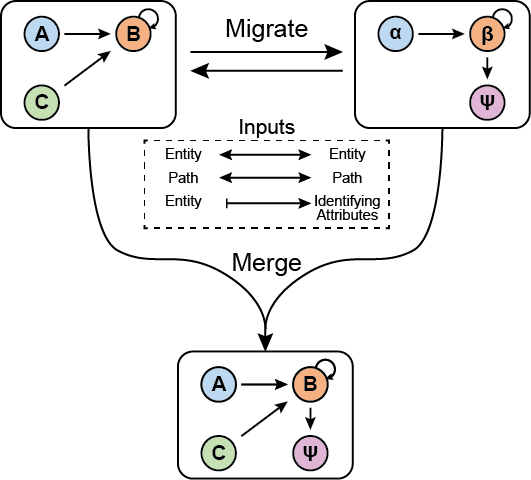}
\caption{Example operations and interface for scientists integrating information from related schemas, mediated by CQL. We color code entities to communicate whether they share the same meaning. Note that \textit{merge} and \textit{migrate} are not primitive operations in CQL, but rather they are the result of a series of commands in CQL.}
\label{sum}
\end{figure}

The data sharing problems described in Section 1.2 strongly apply to computational materials science due to the complexity of representing data in this domain. Thus, the field could benefit from a user-friendly technology (Figure \ref{sum}) that can offer researchers a means of integrating heterogeneous data sources. To demonstrate the flexibility of CQL as a tool for interconversion in materials science, we develop case studies for data integration using the Open Quantum Materials Database (OQMD) below.
\footnote{This perspective will only consider the integration of relational databases, which have a well-defined structure to their data. There are important trade-offs to consider between the relational data model and alternatives (most prominently, MongoDB).  OQMD \cite{oqmd} is the only major materials database that currently is relational, in contrast to the Materials Project\cite{mp}, Inorganic Crystal Structure Database (ICSD) \cite{icsd}, and NOMAD \cite{nomad} databases. 
}

\section{Case Studies}

\subsection{Overview}

We will consider two problems: data migration and data integration. In addition to OQMD, we will consider a representative materials database called Catalysis, with a rich schema structure that encodes information in a different way from OQMD. Considering the perspective of a materials scientist who has represented their work for a particular problem in Catalysis yet wants to incorporate data from OQMD, we \begin{enumerate}
    \item migrate OQMD $ \,\to\,$ Catalysis. The resulting instance has the same schema as Catalysis and data from both databases.
    \item merge OQMD and Catalysis into a new schema that includes overlapping and non-overlapping content of both databases.
\end{enumerate}

Readers are encouraged to consider the example input files and databases and to check references made to code in the footnotes. Below is a summary of questions whose answers are required to incorporate data from an external database into one's local structured data schema:

\begin{itemize}

\item What tables / \textit{paths} correspond to the same meaning? \footnote{\textbf{overlap.py} Section 1.1}

\item What functions, if any, are required to translate between two representations of data with same meaning (e.g.\ unit conversion, formatting)? \footnote{\textbf{javafuncs.py}}

\item Are there any attributes which are not explicitly represented (describable as a constant or a function of other attributes in the database)?   
\footnote{We can can choose whether to compute these attributes in SQL while 'landing' the data into CQL or to compute them within CQL itself (\textbf{overlap.py} Section 3)} 

\item If we only want to consider some subset of records in a database, what filters should be applied?
\footnote{\textbf{oqmd.py} Section 2}

\item What is sufficient for two records to be considered the same despite being represented differently in the two different databases? 
\footnote{Specified when declaring the \textit{identifying} attributes and relations for entities in \textbf{oqmd.py} and \textbf{catalysis.py}. This concerns record linkages, identifying when the same record is referred to in separate databases. As an example: one would not want the number of elements to double when merging together two databases that both contain records for chemical elements.}

\end{itemize}

As it is infeasible to discuss every detail of the three scenarios, it will suffice to describe some \textit{general} sources of differences between the various schemas. We will show specific examples for each of these general problems and how they were addressed. 

\subsection{Factors leading to heterogeneity}

\subsubsection{Differing names}

Two databases could appear different solely because differing names were chosen for entities or attributes which have the same meaning. This mapping of names can be directly specified. \footnote{\textbf{overlap.py} Section 1.2. Special care should be taken to make sure entities with similar names actually do share the same meaning. For example, \textbf{OQMD.structures} and \textbf{Catalysis.structure} both describe an entity consisting in atoms with positions within a crystal lattice; however, \textbf{OQMD.structures} has attributes for \textit{energy} and other DFT-computed properties. This means OQMD structures implicitly have a specific calculation associated with them, whereas Catalysis structures contain purely geometric information (many calculations could share the same structure). Thus, these entities play \textit{structurally} different roles within their respective schemas, despite the similarity their names and many of their attributes. CQL will raise an error during static analysis of the input file if one attempts to unify these objects without first removing certain attributes from \textbf{OQMD.structures}.}

\subsubsection{Implicit constants}

If a dataset never varies a particular parameter, then it is likely to not explicitly include an attribute for that parameter, even if it is critical for the interpretation of the data (the value of that parameter must be communicated through database metadata). This is the case for OQMD, where every calculation in the database uses VASP as the code which implements DFT. When migrating data from OQMD to a database which has the DFT code as a parameter, we can declaratively specify this database-wide constant. \footnote{\textbf{overlap.py} Section 1.3}

\subsubsection{Degree of Denormalization}

A normalized database schema groups together repeating co-occurrences of data into refined structures. This is often a design trade-off between ease of reading from and writing to the database \cite{denorm}. For example, every chemical structure exists within a periodic crystal unit cell, representable with three vectors in $\mathbb{R}^3$ (encoded as nine floating point numbers). If many atomic structures share the same unit cell, then duplication can be avoided by representing the cell as a distinct entity and giving \textbf{Structure} a foreign key to \textbf{Cell}. This strategy is taken by Catalysis; however, in OQMD the cell properties are directly found as attributes to \textbf{OQMD.structures}. The solution is to map attributes of \textbf{OQMD.structures} to paths in Catalysis. 
\footnote{\textbf{overlap.py} Section 1.2}

\begin{figure}[htp]
\centering
\includegraphics[width=14cm]{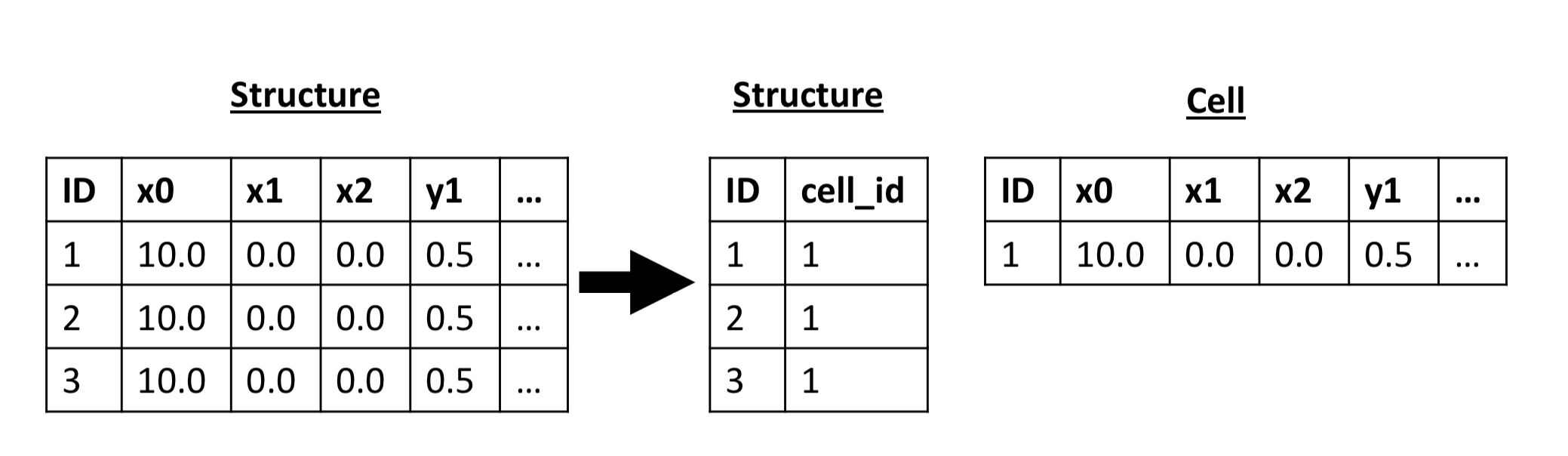}
\caption{Example variation due to normalization. A path mapping required for this migration is $Structures\ .\ x0\ \mapsto Structure\ .\ cell\_id\ .\ x0$, among others.}
\label{denorm.png}
\end{figure}

\subsubsection{Hidden structure}

Attributes sometimes store data with internal structure, in effect hiding the structure from the schema. This can pose a challenge for specifying the relation between two schemas by path-based means alone. However, CQL allows for  functions to be integrated into path equations to address the specific ways in which data should be extracted from an attribute. For example, OQMD has, for each DFT calculation, an attribute \textit{params}, which is a JSON object functioning as an implicit tuple. Conversely, these attributes are explicit attributes for each DFT calculation in Catalysis, and migrating the information requires specifying a function which can unpack the JSON and extract the correct field. \footnote{\textbf{overlap.py} Section 1.3 }

\subsubsection{Different levels of granularity}

Every model at some point has to make a cutoff about what level of detail should be recorded, thus sometimes features of a target schema are not present in the source schema, nor is there sufficient information to compute the missing information. This can be addressed by leaving attributes as \textbf{NULL}\footnote{In contrast to SQL, CQL uses \textit{labelled} nulls, which are distinct (but unknown) values that behave like variables. This allows for unknown	data to	be	manipulated	as	if	they	were	actual	values, participating in user-defined functions and data-integrity constraints.} (or choosing some value that communicates a lack of knowledge) when migrating from low granularity to high granularity, whereas it is trivial to ignore excess attributes and foreign keys when migrating data in the other direction.

\section{Future Outlook}

\begin{figure}[htp]
\centering
\includegraphics[width=14cm]{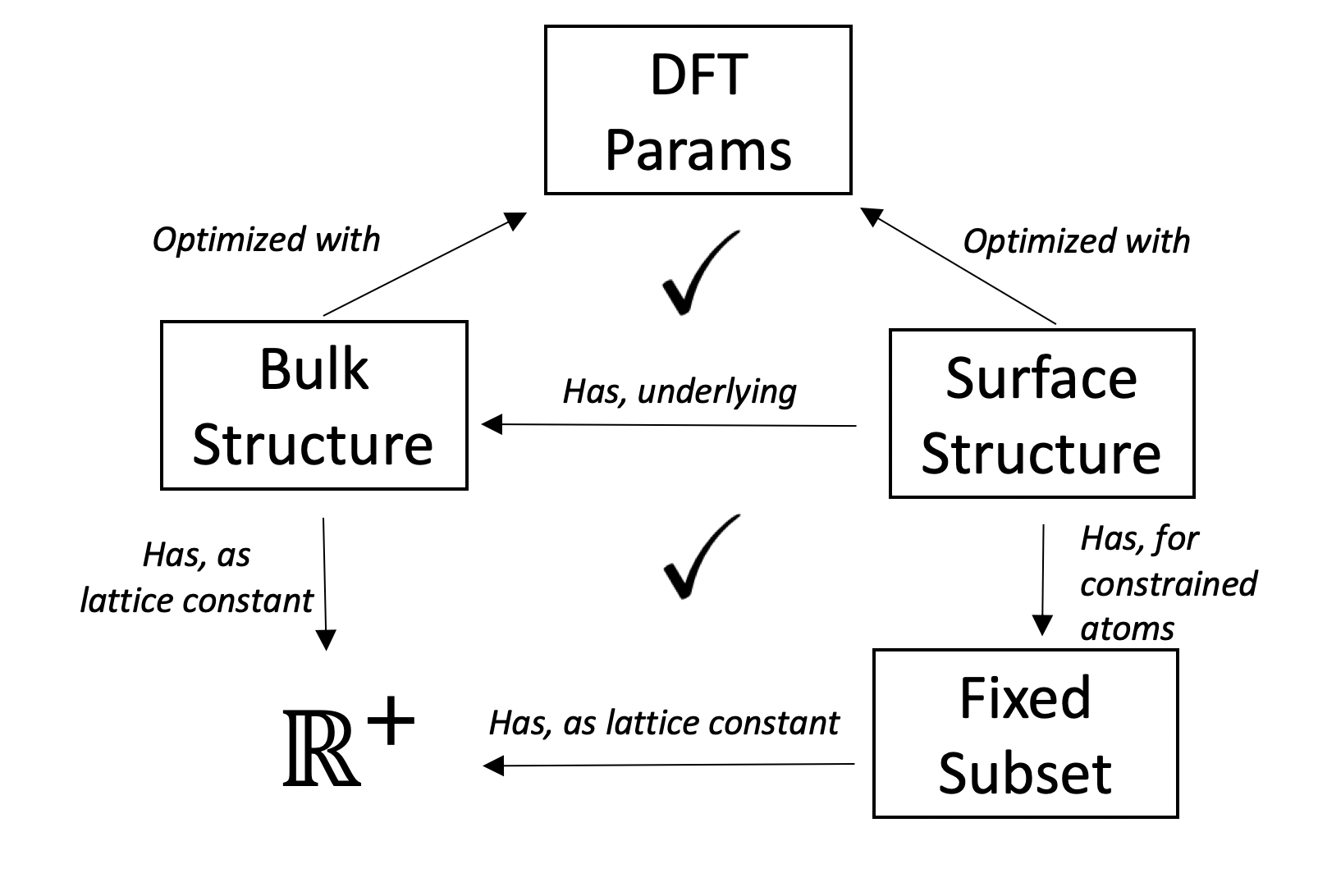}
\caption{Example data integrity diagram for surface chemistry structures, where for any instance data to be sensible we would demand path equalities for the top triangle and the bottom rectangle. $\mathbb{R}^+$ represents positive real numbers. One can imagine this diagram being integrated with that of Figure \ref{olog}, sharing the \textbf{Bulk Structure} and \textbf{Surface Structure} entities.}
\label{surf}
\end{figure}

The example databases from this study only considered bulk materials, which are comparatively simple relative to surfaces. Surface DFT models have many practical applications yet also involve more complicated relations between calculations and structures. We plan to demonstrate the utility of the algebraic data integration with CQL in surface science (Figure \ref{surf}). Considering a broader outlook, this perspective focused on materials science and DFT, yet the tools are equally applicable to other scientific fields. For example, life sciences research broadly speaking lacks widespread data standardization, and survey results have identified and urgent need for "user-friendly tools targeting integration of heterogeneous datasets"\cite{omics}. Lastly, CQL is an expressive language for performing a large variety of database manipulations, of which we have only considered a particular kind of migration and a particular kind of merging. 

\section{Conclusion}

The lack of a robust method for moving data between complex schemas has discouraged the sharing of data. We demonstrate that CQL offers an intuitive language for precisely specifying the translation between different structured data representations, thus expanding the usefulness of structured data and facilitating precise communication between researchers with different frames of reference. This has been demonstrated with data migration and data integration examples for a real-world problem from computational materials science.

Open problems in data sharing have been summarized as the lack of 1.) structured data, 2.) tools to combine information from disparate datasets, and 3.) communication of the nuances of data sets. Structure-preserving data integration operations offer a means of addressing these needs while allowing for individualized data models. 

\section{Declaration of Interest}

KB was supported by the Department of Defense (DoD) through the National Defense Science \& Engineering Graduate Fellowship (NDSEG) Program. RW and DS are consulting members for Categorical Informatics, which offers consultation services for CQL usage.

The authors would like to thank Michael Statt from Stanford University and Joseph Montoya from Toyota Research Institute for constructive criticism of the manuscript.

\section{Data Availability}

Data and code required to reproduce these findings are available to download from https://github.com/kris-brown/cql\_data\_integration.

\bibliographystyle{alpha}

\pagebreak 

\bibliography{sample}

\end{document}